\documentstyle[prl,aps,floats]{revtex}
\begin{document}
\renewcommand{\topfraction}{0.8}
\twocolumn[\hsize\textwidth\columnwidth\hsize\csname
@twocolumnfalse\endcsname
\begin{center}
{\Large \bf Effective time variation of $G$ in a model}\\ 
\vspace{0.2 cm}
{\Large \bf universe with variable space dimension}\\
\vspace{0.3 cm}
{\large Reza Mansouri$^{a,b}$, Forough Nasseri$^{a,b}$,
Mohammad Khorrami$^{a,c}$}\\
\vspace{0.2 cm}
{\small \it $^{a}$Institute for Studies in Theoretical Physics and Mathematics,
P.O.Box 19395--5531, Tehran, Iran\\
$^{b}$Department of Physics, Sharif University of Technology, P.O.Box
11365--9161, Tehran, Iran\\
$^{c}$Institute for Advanced Studies in Basic Science, P. O. Box, 45195--159,
Gava Zang, Zanjan, Iran\\}
\end{center}
\begin{abstract}                                                     
---------------------------------------------------------------------------------------------------------------------------------

{\bf Abstract}\\
Time variation of Newtonian gravitational constant, $G$,
is studied in model universes with variable space dimension proposed
recently. Using the Lagrangian formulation of these models, 
we find the effective gravitational constant as a function of time. To
compare it with observational
data, a test theory for the time variation of $G$ is formulated. We have
assumed a power law behavior of the time variation of $G$ where the exponent
$\beta$ is itself time dependent. Within this test theory we are able to
restrict the free parameter of the theories under consideration and give
upper bounds for the space dimension at the Planck era.
The time variation of $G$ at earlier times, such as the time of
nucleosynthesis is also predicted which express the needs to look for
related observational data. \\

---------------------------------------------------------------------------------------------------------------------------------
\end{abstract}
\vskip2pc]
{\bf 1. Introduction}\\
The present value of Newtonian gravitational constant, $G$, is known with 
the least accuracy \cite{1,2}. There are three problems connected with $G$:
absolute value measurements and possible variations with time \cite{3,4} and 
space \cite{5,6}. There is a promising new space experiment SEE--Satellite
Energy Exchange--\cite{7} which addresses to all these problems and may
be more 
effective in solving them than other laboratory or space experiments. The 
SEE project may improve our knowledge of $G$, limits on temporal and space 
variation by $2-3$ orders of magnitude \cite{8}.\\
Current ideas suggest that the value of $G$ might be related to other 
fundamental constants of physics \cite{9}. It has been recently suggested 
that gravity could be even more drastically modified below some distance,
$r_0$. In principle, Cavendish--type experiments performed for separation
smaller than $r_0$ might see a change of the $1/{r^2}$ law: the exponent $2$ 
being replaced by an exponent larger than or equal to $4$.
In these models, $r_0$ is already constrained to be smaller than 
$\sim 1 \mu m$ \cite{10}. $G$--measurements and Cavendish--type experiments 
have now reached a new significance as possible windows on the physics of 
unification between gravity and the other interactions.\\
It must be noted that the variability of $G$ is explicitly or effectively
model dependent. In several modified quantum theories of gravity, scaling of
$G$ with the distance is standard \cite{11,12,13,14}. However, the
modification of $G$ is small and Newtonian gravity holds in the weak energy
limit. Confirmations of this scheme are coming from satellites' measurements
of long acceleration \cite{15}.\\
Interest in the problem of time variation of $G$ has increased greatly 
during the last decade because of new developments in Kaluza--Klein and 
superstring theories of the unification of all physical interactions.
Observational bounds on the time evolution of extra spatial dimensions
in Kaluza--Klein and superstring theories can be obtained from limits on 
possible variation of $G$ and other constants \cite{16,17,18,19}.
The recent version of the dilaton evolution proposed by Damour and 
Polyakov \cite{20} in the context of string theory has provided an 
expression connecting the time variation of $G$ and that of fine structure 
constant, $\alpha$. Barrow \cite{21} has discussed the variation of $G$ in 
Newtonian and relativistic scalar--tensor gravity theories.\\
In this paper, we will discuss on the effective time variation of $G$ in  
the models proposed in \cite{22,23,25}.
In these models, the space dimension varies with the expansion of the
universe. The model proposed in \cite{22,23} seems to be singularity free,
having two turning points for the space dimension. This model
has been criticized in \cite{24}. The way of generalizing the standard 
cosmological model to arbitrary space dimension used in \cite{22,23} is
questioned, and another way of writing the field equations is proposed.
It has been pointed out that the model given in \cite{22,23} has no upper
bound for the dimension of space.\\
Later on, we studied critically the previous works in \cite{22,23,24},
and derived new Lagrangians and field equations. We also discussed the model
universe with variable space dimension from the view point of quantum
cosmology and obtained a general wave function for this model. In the limit
of constant space dimension, our wave function approaches the tunneling
Vilenkin wave function or the modified Linde wave function \cite{25}.\\
Here we are interested to study the time variation of $G$ within these
cosmological models with variable space dimension. There is a hope that
observational data for time variation of $G$ may distinguish between these
theories or show their viability.\\
Every comparison of a theory with experiment or observation needs a socalled
test theory (for examples in special relativity see \cite{26}). Such a 
formulation for $G$ variation is given in \cite{27}, which is however not 
general enough to include models we are interested in. Therefore a new 
generalized test theory for time variation of $G$ is formulated in section 2. 
A short review of the cosmological models with variable space dimension is 
given in section 3. Based on the test theory of section 2, we are able to 
restrict the free parameter of the theories under consideration to be
consistent with observation. This gives us upper bounds for the dimension
of space at the Planck era. Our generalized test theory allows to predict
the variation of $G$ at even earliest times. Therefore, the value of $G$
at the time of nucleosynthesis can be compared to that of the present
ime.\\
\vspace{0.3 cm}\\
{\bf 2. Formulation of a one parameter test theory}\\
{\bf for the time variation of $G$}\\
\vspace{0.1 cm}\\
It is always very fruitful to have a theory for any tests of physical theories
or constants (see for example \cite {26} for an extensive use of test
theories in special relativity). Test theories are not only helpful to
interpret the experimental or observational data but they also direct us to
new and sometimes crucial tests. Therefore one should care about their
formulations. Here we will elaborate on an existing formulation for the
time variation of $G$ \cite{27} and generalize it to more physical cases.
Let us take
\begin{equation}
\label{14}
G(t)=G_0 (\frac{t}{t_0})^{\beta},
\end{equation}
where $t_0 \simeq 10^{17} {\rm sec}$ is the present time and 
$G_0$ is the present value of $G$. 
In this case, the average rate of the cosmological time variation of  
$G$ is
\begin{equation}
\label{15}
\frac{\dot G}{G}= \frac{\beta}{t},
\end{equation}
so that today
\begin{equation}
\label{16}
{\left( \frac{\dot G}{G} \right) }_0=\frac{\beta}{t_0}.
\end{equation}
Data from big bang nucleosynthesis yields
\begin{equation}
\label{17}
|\beta| \lesssim 0.01,
\end{equation}
or 
$
({\dot G}/{G})\lesssim 10^{-12} {\rm yr}^{-1}
$ \cite{27}.
We now generalize the relation (\ref {14}) to the cases where $\beta$ is not
constant but a function of time:
\begin{equation}
\label{30}
G=G_0(\frac{t}{t_0})^{\beta(t)}.
\end{equation}
Time derivative of this equation yields
\begin{equation}
\label{31}
\frac{\dot G}{G}=\frac{\beta(t)}{t}+{\dot {\beta}}(t)
\ln (\frac{t}{t_0}).
\end{equation}
Now, if $\beta (t)$ and its time derivative, ${\dot \beta}(t)$,
satisfy the following condition
\begin{equation}
\label{32}
|\frac{{\dot {\beta}}(t)}{\beta(t)}| \ll 
|\frac{1}{t \ln \frac{t}{t_0}}|,
\end{equation}
Eq.(\ref{31}) can be written as
\begin{equation}
\label{33}
\frac{\dot G}{G} \simeq \frac{\beta (t)}{t},
\end{equation}
which looks like (\ref {16}) with time dependent $\beta$. The condition
(\ref {32}) may not always be valid. Therefore it must be checked for each
case. This test theory shows for example how time variation of $G$ could be
different in different eras. We are then led to find observation giving
data for time variation of $G$ in early universe or later times. We will see
in the next section that models with variable space dimensions can be handled
within this new test theory.\\
\vspace{0.3 cm}\\
{\bf 3. Review of the model universe with variable}\\
{\bf space dimension}\\ 
\vspace{0.1 cm}\\
The model introduced in \cite{22,23} is based on a flat Friedmann universe
with dynamical space dimension. Here we do not restrict ourselves to a 
special topology and introduce $k=-1, 0, +1$ for the open, flat, or closed
model, respectively. The Lagrangian of our model universe is
\begin{eqnarray}
\label{1}
{\cal L}&:=&-\frac{D(D-1)}{2 \kappa N} [(\frac{\dot a}{a})^2 -\frac{N^2 
k }{a^2}](\frac{a}{a_0})^D+\frac{1}{2} (-\hat \rho N^2 \nonumber\\
&+&\hat p D a^2),
\end{eqnarray}
where
$$
\hat \rho:=\frac{rho}{N}(\frac{a}{a_0})^D,
\hat p:= p a^{-2} N (\frac{a}{a_0})^D.
$$
There is a constraint in this model which can be written as 
\begin{equation}
\label{2}
(\frac{a}{\delta})^D= (\frac{a_0}{\delta})^{D_0}= e ^{C},
\end{equation}
or
\begin{equation}
\label{3}
\frac{1}{D}=\frac{1}{C} \ln (\frac{a}{a_0})+\frac{1}{D_0}.
\end{equation}
Here is $a$ the scale factor of the Friedmann universe, $N$ the lapse
function, $D$ the variable space dimension, $\rho$ the energy density, $p$
the pressure, $\delta$ the characteristic minimum length of the model, $C$
a constant of the model, and $\kappa=8 \pi G$. The zero subscript in any
quantity, e.g. in $a_0$ and $D_0$, denotes its present value.
Note that in Eqs.(\ref{1}, \ref{2}), the space dimension is a function of 
cosmic time, $t$. Time derivative of Eq.(\ref{2}) leads to 
\begin{equation}
\label{4}
\dot D= -\frac{{D^2}{\dot a}}{C a}.
\end{equation}
It is easily seen that the case of constant space dimension corresponds
to the limit of $C \to +\infty$. Varying the action (\ref{1}) with respect to $N$ and $a$, 
and taking $\hat \rho$ and $\hat p$ as constant, we arrive at the following 
equations of motion in the gauge $\dot N =0$
\begin{equation}
\label{5}
\frac{1}{N^2} (\frac{\dot a}{a})^2 +\frac{k}{a^2} =\frac{2 
\kappa \rho}{D(D-1)},
\end{equation}
and
\begin{eqnarray}
\label{6}
&&\frac{\ddot a}{a}+[\frac{D^2}{2 D_0}-1-\frac{D(2D-1)}{2C(D-1)}]
\{ (\frac{\dot a}{a})^2 +\frac{N^2 k}{a^2} \} \nonumber\\
&& +N^2 \kappa p (\frac{1- \frac{D}{2C}}{D-1})=0.
\end{eqnarray}
Using these equations of motion, one can easily obtain the 
continuity equation
\begin{equation}
\label{con}
\frac{d}{dt}(\rho (\frac{a}{a_0})^D)+p \frac{d}{dt} (\frac{a}{a_0})^D=0.
\end{equation}
In the limit of constant space dimension, or $C \to +\infty$, Eqs. (\ref{1},
\ref{5}, \ref{6}, \ref{con}) approach to the corresponding equations
for constant space dimension $D=D_0$: 
\begin{eqnarray}
\label{7}
&&{\cal L}^0:=
\{ -\frac{D_0(D_0 -1)}{2 {\kappa}_0 N} [(\frac{\dot a}{a})^2 
-\frac{N^2 k}{a^2}] \nonumber\\
&&+\frac{N}{2}(-\rho +Dp)\} (\frac{a}{a_0})^{D_0},\\
\label{8}
&&\frac{1}{N^2} (\frac{\dot a}{a})^2 +\frac{k}{a^2} =\frac{2 
\kappa_0 \rho}{D_0(D_0-1)},\\
\label{9}
&&\frac{\ddot a}{a}+(\frac{D_0-2}{2}) \{ (\frac{\dot a}{a})^2 +
\frac{N^2 k}{a^2} \} \nonumber\\
&&+\frac{N^2 \kappa_0 p}{D_0-1}=0,\\
&&\frac{d}{dt}(\rho a^{D_0})+p \frac{d}{dt}(a^{D_0})=0.
\end{eqnarray}
We have introduced $\kappa_0$ for the value of the gravitational coupling
constant in the case of the constant space dimension $D = D_0$.
In Ref. \cite{25}, we have mentioned some of the shortcomings of
the original model proposed in \cite{22,23}, regarding the field 
equations and their results.
There, we have mentioned that the Lagrangian is not unique. Using 
Hawking--Ellis action of a prefect fluid, we have obtained the following
Lagrangians for the model universe with variable space dimension:
\begin{eqnarray}
\label{10}
&&{\cal L}_I:= -\frac{V_D}{2 \kappa} (\frac{a}{a_0})^D \frac{D (D-1)}{N}
\{ (\frac{\dot a}{a})^2 -\frac{N^2  k}{a^2} \} \nonumber\\
&&- \rho N V_D (\frac{a}{a_0})^D,
\end{eqnarray}
and
\begin{eqnarray}
\label{11}
&&{\cal L}_{II}:= - \frac{V_D}{2 \kappa} (\frac{a}{a_0})^D
\{ \frac{2 {\dot D} {\dot a}}{a N}+\frac{2 D {\dot a}{\dot D}}{aN} \ln  
\frac{a}{a_0}+\frac{D(D-1)}{N}\nonumber\\
&& \times \{ (\frac{\dot a}{a})^2 -\frac{N^2 k}{a^2} \}+
\frac{2 D {\dot D}{\dot a}}{Na} \frac{d \ln V_D}{dD} \}\nonumber\\ 
&&- \rho V_D N (\frac{a}{a_0})^D.
\end{eqnarray}
Here $V_D$ is the volume of the spacelike sections:
\begin{eqnarray}
\label{12}
V_D &=& \cases {\frac{2 \pi^{(\frac{D+1}{2})}}{\Gamma(\frac{D+1}{2})},          
	       & if $k=+1$, \cr          
	       \frac{\pi^{(\frac{D}{2})}}{\Gamma(\frac{D}{2}+1)}{\chi_c}^D,          
	       & if $k=0$, \cr
	       \frac{2\pi^{(\frac{D}{2})}}{\Gamma(\frac{D}{2})}f(\chi_c),            
	       & if $k=-1$, \cr}
\end{eqnarray}
where $\chi_c$ is a cut--off and $f(\chi_c)$ is a function thereof 
(see Ref.\cite{25}). In the limit of constant space dimension, 
${\cal L}_I$ and ${\cal L}_{II}$ approach to
\begin{eqnarray}
\label{13}
&&{\cal L}^0_{I,II}:= -\frac{V_{D_0}}{2 \kappa_0 N}(\frac{a}{a_0})^{D_0} 
D_0 (D_0-1)[(\frac{\dot a}{a})^2 -\frac{N^2 k}{a^2}] \nonumber\\
&&- \rho N V_{D_0}(\frac{a}{a_0})^{D_0}.
\end{eqnarray}
A complete discussion of the field equations
corresponding to ${\cal L}_I$ and ${\cal L}_{II}$, the dynamics of the model, 
and the wave function of the universe have been given in Ref.\cite{25}.\\
We define $D_{Pl}$ as the space dimension of the universe when the scale
factor is equal to the Planck length, $l_{Pl}$.
Taking the scale of universe at $D_0=3$, to be the present value of the
Hubble radius, $H_0^{-1}$, and the space dimension in the Planck length
to be $4$, $10$, or $25$, coming from Kaluza--Klein and superstring theories,
we can obtain from Eqs.(\ref{2},\ref{3}), the corresponding value of
$C$ and $\delta$, see Table I. \\
\begin{center} 
\small{TABLE I. Values of $C$ and $\delta$ for some interesting\\ 
values of $D_{Pl}$ assuming $D_0=3$.\\
}
\begin{tabular}{|p{1.0 cm}|p{1.0 cm}|p{2.0 cm}|}  \hline\hline
 $C$ & $D_{pl}$       & $\delta ({\rm cm})$ \\ \hline
 $\infty$ & $3$       & $0$ \\  \hline
 $1678.8$ & $4$       & $8.6 \times 10^{-216}$ \\  \hline
 $599.57$ & $10$      & $1.5 \times 10^{-59}$ \\  \hline                
 $476.93$ & $25$      & $8.4 \times 10^{-42}$ \\  \hline
 $419.70$ & $+\infty$ & $l_{Pl}$ \\ \hline\hline
\end{tabular}
\end{center}
\vspace{0.2 cm}
{\bf 4. Time variation of $G$ in model universe with variable
space dimension}\\
\vspace{0.1 cm}\\   
Take first the model based on $\cal L$. The gravitational coupling constant
$\kappa$ have to be related to $G$. Although this relation seems not to be
trivial \cite {28}, we will assume for simplicity the familiar relation
$\kappa = 8 \pi G$. Now, comparing the coefficients of $({\dot a}/{a})^2$
in ${\cal L}^0$ and ${\cal L}$ we obtain
\begin{equation}
\label{18}
\frac{G}{D(D-1)}= \frac{G_0}{D_0(D_0-1)}.
\end{equation}
This may also be written as
\begin{equation}
\label{19}
G=G_0 f(D),
\end{equation}
where
\begin{equation}
\label{20}
f(D):=\frac{D(D-1)}{D_0(D_0-1)}.
\end{equation}
Time derivative of Eq.(\ref{19}) yields
\begin{equation}
\label{21}
\frac{\dot G}{G}= {\dot D} \frac{f'(D)}{f(D)} =
-\frac{D^2 {\dot a} {f'}(D)}{C a f(D)},
\end{equation}
where prime denotes the derivative with respect to $D$. To
express the time evolution of $G$ given by Eq.(\ref{21}) in
terms of $t$, we have to know $a$ and $D$ as functions of
the time. This requires the solution of the equation of
motion (\ref{5}, \ref{6}) which does not seem to have any
analytic solution. As an approximation, we take the solution
of the standard big bang cosmology in the radiation dominated
(RD) and matter dominated (MD) universe for the constant $D$
space dimension: 
\begin{eqnarray}
\label{22}
a=a_0 (\frac{t}{t_0})^{1/2},\,\Rightarrow a=a_0(\frac{t}{t_0})^{2/(D+1)},\,
\,\,{\rm RD}\,\,{\rm universe}, \\
\label{23}
a=a_0 (\frac{t}{t_0})^{2/3},\,\Rightarrow a=a_0(\frac{t}{t_0})^{2/D},\,
\,\,{\rm MD}\,\,{\rm universe}. 
\end{eqnarray}
Assuming now $D$ to be variable, and using (\ref{4}), the time 
derivative of Eqs.(\ref{22}) and (\ref{23}) yield,
respectively:
\begin{eqnarray}
\label{24}
\frac{\dot a}{a}&=&\frac{2}{t[1+\frac{D^2}{D_0}]},
\,\,\,\,\,{\rm RD}\,\,\,{\rm universe},\\
\label{25}
\frac{\dot a}{a}&=&\frac{2 D_0}{D^2 t}, \,\,\,\,\,\,\,\,{\rm MD}
\,\,\,{\rm universe}.
\end{eqnarray}
Inserting (\ref{24}, \ref{25}) in (\ref{21}), we obtain the
time evolution of $G$ for the radiation--dominated universe
\begin{equation}
\label{26}
\frac{\dot G}{G} \simeq \frac{\beta_{RD}}{t},
\end{equation}
and for the matter--dominated universe
\begin{equation}
\label{27}
\frac{\dot G}{G} \simeq \frac{\beta_{MD}}{t},
\end{equation}
where
\begin{eqnarray}
\label{28}
\beta_{RD}&=& -\frac{2 D^2 f'(D)}{C[1+\frac{D^2}{D_0}]f(D)},\\
\label{29}
\beta_{MD}&=& -\frac{2 D_0 f'(D)}{C f(D)}.
\end{eqnarray}
We may take $\beta_{RD}$ or $\beta_{MD}$ given by (\ref{28}) or (\ref{29})
for $\beta(t)$ in Eqs.(\ref{31},\ref{32},\ref{33}).
But first notice that for the present time
$t_0$, the second term in the $RHS$ of Eq.(\ref{31}) vanishes and
Eq.(\ref{33}) is exactly true. The present value of $\beta(t)$, say
$\beta_{0}$, is obtained by inserting $D=D_0=3$ in Eq.(\ref{29}). Note that
${\beta}_0$ depends on the $C$--parameter and consequently on $D_{Pl}$.
Table II shows some values of $\beta_0$.\\
The earliest time for which we have observational data is the time of
nucleosynthesis, $t_{ns} \sim 1 {\rm sec}$. Using the standard big bang
cosmology, the scale factor of the universe at the
nucleosynthesis time is nearly $a_{ns} \sim 3 \times
10^{18} {\rm cm}$. The corresponding value of the
space dimension, $D_{ns}$, can be obtained by inserting
$a=a_{ns}$, $a_0=H_0^{-1} \simeq 10^{28} {\rm cm}$ and $D_0=3$ in
Eq.(\ref{3}). For each value of $C$ or $D_{Pl}$, the
corresponding value of $D_{ns}$ is given in Table II.
The value of $\beta$ at the time of nucleosynthesis,
$\beta_{ns}$, is now obtained from (\ref {28}) using the
appropriate $D_{ns}$ value. It can now be easily seen that
the condition (\ref {32}) is still valid for $t=t_{ns}$.
Therefore,  Eq.(\ref {33}) can be used for the time variation
of $G$ at $t_{ns}$. Numerical values for $\beta_{ns}$ and
${G_{ns}}/{G_0}$ are obtained for some values of $D_{Pl}$
in Table II. We are not aware of any observational data relating
to this quantity, but given observational data one can look for the
validity of the underlined theory.\\ 
Let us now compare the numerical value of $\beta_{0}$ for the model $\cal L$
with observational data, $|\beta| \lesssim 0.01$, see Eq.(\ref{17}). From
Table II, we see that for $D_{Pl} > 25$, the corresponding absolute values
of $\beta_{0}$ are bigger than $0.01$. Hence, models defined by $\cal L$
having $D_{Pl} > 25$ are ruled out.\\
Let us now consider the Lagrangians ${\cal L}_I$ and
${\cal L}_{II}$. In this case, the comparison of the coefficients of 
$({\dot a}/{a})^2$ in ${\cal L}_{I}$ and ${\cal L}_{II}$ 
with ${\cal L}^0_{I,II}$ give us the relation between $G_0$ and $G$:
\begin{equation}
\label{34}
\frac{G}{V_D D(D-1)}=\frac{G_0}{V_{D_0} D_0 (D_0-1)},
\end{equation}
which can be written as
\begin{equation}
\label{35}
G=F(D) G_0,
\end{equation}
where
\begin{equation}
\label{36}
F(D):=\frac{V_D D(D-1)}{V_{D_0} D_0 (D_0-1)}.
\end{equation}
For simplicity, let us now assume a closed Fridmann universe with $k = +1$.
From Eq.(\ref{36}) we then obtain
\begin{equation}
\label{37}
F(D)=\frac{\pi^{D/2} D(D-1) \Gamma (\frac{D_0+1}{2})}
{\pi^{{D_0}/2} D_0(D_0-1) \Gamma (\frac{D+1}{2})}.
\end{equation}
Similar to our treatment for ${\cal L}$, it can be shown that for the
Lagrangians ${\cal L}_I$ and ${\cal L}_{II}$, the time evolution of $G$ is given
by Eq.(\ref{30}), where  $\beta$ is defined according to (\ref{28},\ref{29}) except
that $f(D)$ is replaced by $F(D)$. It is also seen that the condition
(\ref{32}) is still valid for the time of nucleosynthesis. Some interesting
values for $\beta_0$, and $\beta_{ns}$, and ${G_{ns}}/{G_0}$ are given in
Table III. \\
Comparing the values of $\beta_{0}$ from Table III with observational
data, given by Eq.(\ref{17}), models with $D_{Pl} = 10, 25$, and so are 
ruled out.\\
\vspace{0.25 cm}  
\begin{center}
\small{TABLE II. Values of $C$, $D_{Pl}$, $D_{ns}$, $\beta_0$,
$\beta_{ns}$, $G_{ns}/G_{0}$,\\
based on ${\cal L}$ and ${\cal L}^{0}$. The limit of $C \to +\infty$\\
corresponds to the constant space dimension.\\
}
\begin{tabular}{|p{1.0 cm}|p{0.75 cm}|p{0.8 cm}|p{1.25 cm}|p{1.25 cm}|p{1.25
cm}|}  \hline\hline
 $C$       & $D_{Pl}$  & $D_{ns}$ & $\beta_0$ & $\beta_{ns}$ & $G_{ns}/G_0$ \\ \hline
 $+\infty$ & $3$       & $3$      & $0$       & $0$          & $1$    \\ \hline 
 $1678.8$  & $4$       & $3.12$   & $-0.003$  & $-0.0021$    & $1.09$ \\ \hline 
 $599.57$  & $10$      & $3.37$   & $-0.008$  & $-0.0057$    & $1.26$ \\ \hline
 $476.93$  & $25$      & $3.48$   & $-0.010$  & $-0.0070$    & $1.32$ \\ \hline
 $419.70$  & $+\infty$ & $3.56$   & $-0.012$  & $-0.0078$    & $1.33$ \\ \hline\hline
\end{tabular}
\end{center}
\vspace{0.25 cm}  
\begin{center}
\small{TABLE III. Values of $C$, $D_{Pl}$, $\beta_0$, $\beta_{ns}$,
$G_{ns}/G_0$ based\\
on ${\cal L}_I$, ${\cal L}_{II}$, and ${\cal L}^{0}_{I,II}$.
}
\begin{tabular}{|p{1.0 cm}|p{0.75 cm}|p{1.25 cm}|p{1.25 cm}|p{1.25 cm}|}
\hline\hline
 $C$       & $D_{Pl}$  & $\beta_0$ & $\beta_{ns}$ & $G_{ns}/G_0$ \\ \hline
 $+\infty$ & $3$       & $0$       & $0$          & $1$    \\ \hline 
 $1678.8$  & $4$       & $-0.004$  & $-0.003$     & $1.13$ \\ \hline 
 $599.57$  & $10$      & $-0.012$  & $-0.008$     & $1.38$ \\ \hline
 $476.93$  & $25$      & $-0.015$  & $-0.010$     & $1.49$ \\ \hline
 $419.70$  & $+\infty$ & $-0.017$  & $-0.011$     & $1.50$ \\ \hline\hline
\end{tabular}
\end{center}
{\bf 5. Conclusions}\\
\vspace{0.1 cm}\\
The time evolution of $G$ in model universes with variable space dimension
has been studied. Assuming a power law behavior for the time variation of
$G$, it turns out that the exponent $\beta$ has to be time dependent to
generalize enough to include 
models with variable space dimension. This
has led us to a generalized test theory for the time variation of $G$. 
Within this new test theory of $G$ variation, we have shown that theories
based on $\cal L$ are observationally viable for $D_{Pl} \leq 25$. Those
theories based on ${\cal L}_{I}$ and ${\cal L}_{II}$ are observationally 
ruled out for $D_{Pl} \geq 10$. \\
Within this generalized test theory, it is possible to predict the time
variation of $G$ at the era of nucleosynthesis for which it is possible to
have observational data. Therefore, we suggest to look for differences in the
value of $G$ at the present time and the time of nucleosynthesis,
and compare it to the prediction of theories within this test theory.

\end{document}